\pdfoutput=1

\documentclass[reprint, amsmath, amssymb, aps, prd, floatfix]{revtex4-2}

\usepackage[T1]{fontenc}
\usepackage{mathptmx}
\usepackage{graphicx}
\usepackage{dcolumn}
\usepackage{bm}
\usepackage{subcaption}
\usepackage{microtype}
\usepackage[colorlinks, citecolor=blue, urlcolor=blue, linkcolor=blue]{hyperref}

\DeclareCaptionJustification{fulljustified}{\leftskip=0pt\rightskip=0pt\parfillskip=0pt plus 1fil\relax}
\AtEndDocument{\nocite{fullnamesControl}}

\begin{document}
\raggedbottom

\title{Denoising graph superresolution toward improved collider event reconstruction}

\author{Nilotpal Kakati}
\email[Contact author: ]{nilotpal.kakati@weizmann.ac.il}
\author{Etienne Dreyer}
\author{Eilam Gross}
\affiliation{Weizmann Institute of Science, Rehovot, Israel}

\begin{abstract}
In preparation for Higgs factories and energy-frontier facilities, future colliders are moving toward high-granularity calorimeters to improve reconstruction quality. However, the cost and construction complexity of such detectors is substantial, making software-based approaches like superresolution an attractive complementary tool. This study explores integrating superresolution techniques into an LHC-like reconstruction pipeline to effectively enhance calorimeter granularity and suppress noise. We find that this software preprocessing step significantly improves reconstruction quality at fixed detector granularity, narrowing the performance gap to a finer calorimeter without closing it. To demonstrate its impact, we propose a novel transformer-based particle-flow model that offers improved particle reconstruction quality and interpretability. Our results serve as a proof of concept, demonstrating the potential of superresolution to enhance reconstruction in collider environments.
\end{abstract}

\maketitle

\section{Introduction} \label{sec:introduction}
The energy and precision frontiers of high-energy physics are driven by technological advances both in generating particle beams and in recording and reconstructing particle collisions. Detectors at current and future collider facilities are designed to extract precise information about outgoing particles using extensive arrays of sensors, including calorimeter layers. Calorimeters measure energy by instigating a shower of secondary particles, distributing and absorbing the incident particle's energy as it propagates through successive layers of dense material. To help determine particle direction and type, as well as separate nearby particles, calorimeters are segmented into a gridlike array of cells.

Although calorimeter performance is generally characterized by energy resolution, spatial resolution (inversely related to granularity) assumes a crucial role in dense collision environments. At the Large Hadron Collider (LHC)~\cite{lhcMachine} proton-proton collisions yield dense environments rich in hadronic jets and numerous simultaneous collisions (pileup), with a nearly tenfold increase expected in future high-luminosity runs~\cite{ZurbanoFernandez:2020cco}. To reconstruct jet properties accurately, particle-flow (PFlow) algorithms~\cite{ATLASparticleFlow, CMSparticleFlow} leverage tracking information in tandem with calorimeter and other measurements. However, low-granularity calorimeters have limited imaging capabilities, constraining the performance of pattern reconstruction in PFlow algorithms such as assigning energy deposits to tracks, identifying neutral hadrons, and employing software compensation techniques~\cite{PandoraSoftwareCompensation}. These challenges have positioned \textit{particle-flow calorimetry}~\cite{Thomson:2009rp} as a leading design paradigm for future detectors, enabling the 3\% jet energy resolution required at future electron-positron Higgs factories to distinguish hadronic $W$ and $Z$ boson decays~\cite{Aleksa:2021ztd}.

In practical applications, calorimeter granularity is constrained by cost as well as mechanical and electrical engineering challenges, leading to less than optimal granularity. Superresolution (SR) techniques, which have been extensively studied in the field of image processing~\cite{7115171, 7780551, 8014885, 8099502, 10.1007/978-3-030-01234-2_18, 10.1007/978-3-030-11021-5_5, 9887996}, offer the potential to mitigate the limitations of the spatial resolution intrinsic to the calorimeter design by imposing physical priors learned from simulation. By reconstructing high-resolution (HR) data from low-resolution (LR) inputs, these techniques could, in essence, recover substructure features obscured by coarse segmentation, thereby maximizing the performance of a given detector design.

In this study, we explore the integration of superresolution techniques within the core pipeline of particle reconstruction in collider experiments. The initial stage of any collider experiment involves reconstructing low-level particles or particlelike objects, referred to as PFlow objects. By applying superresolution to calorimeter data, we aim to enhance the accuracy of PFlow objects and, thereby, that of higher-level objects used in physics analyses, such as jets, leptons, and photons. We anticipate our approach to be especially useful for reconstructing boosted topologies such as top jets, boosted Higgs decays, and hadronically decaying tau leptons.

\subsection{Related work}

Calorimeter SR was first explored in~\cite{DiBello2021}, which demonstrated a proof of concept using a neural network-based method with a simplified detector and physics setup. Since then, SR has found various applications in particle physics, such as enhancing jet constituent images with generative adversarial networks for improved jet property estimation~\cite{10.21468/SciPostPhys.13.3.064}, employing normalizing flows for upsampling calorimeter showers for fast simulation~\cite{PhysRevD.109.092009}, and producing high-resolution photon shower images~\cite{Erdmann2023}. In the broader context, a large number of generative models have been developed for fast calorimeter shower simulation (see~\cite{Krause:2024avx} for a review). Recent work~\cite{yu2024enhancingeventsneutrinotelescopes} also highlights the potential of SR for enhancing event reconstruction in neutrino telescopes.

\subsection{Key contributions}

Our approach offers two enhancements to calorimeter data to benefit the downstream reconstruction task. The first is the suppression of electronic noise that distorts recorded energy deposits, which we perform by training our model to denoise the energy in each cell. Second, our approach improves spatial resolution by predicting a high-granularity graph of calorimeter cells, allowing the reconstruction of finer details such as nearby particles. Representing the calorimeter data as a graph is a natural choice for handling irregular geometry, nontrivial spatial correlations, and the sparsity in the recorded signal. Superresolution on graph data, where the number of nodes gets upscaled, is not an entirely new concept~\cite{graphSuperTemporal,graphSuperBrain} but is uncommon in the literature. To assess the improvement in reconstruction, we introduce a novel transformer-based PFlow model, based on predicting hypergraphs, as introduced in~\cite{pflow}. This model benefits from modernized architecture and provides improved interpretability, allowing the impact of superresolution to be analyzed transparently.

We summarize our contributions as follows:
\begin{enumerate}
    \renewcommand{\labelenumi}{(\roman{enumi})}
    \setlength{\itemsep}{0pt}\setlength{\parsep}{0pt}
    \item demonstrating superresolution with a full particle reconstruction chain,
    \item denoising energy deposits,
    \item graph-based architecture, and
    \item novel transformer-based particle-flow algorithm.
\end{enumerate}

\section{Datasets} \label{sec:datasets}
To train the superresolution model, we require the same calorimeter shower data in both low and high granularity (i.e., resolution). This was achieved using the \textsc{cocoa} package~\cite{COCOA}, which provides an LHC-like generic calorimeter model simulated with Geant4~\cite{geant1, geant2, geant3}. \textsc{cocoa} includes three electromagnetic (ECAL) and three hadronic (HCAL) concentric layers, each with configurable material, granularity and noise levels. The package emulates sampling fractions of $f=0.07$ and $f=0.025$ in the ECAL and HCAL, respectively, by using an effective molecule that represents the mixture of active (liquid Ar and plastic scintillator) and passive (Pb and Fe) materials. This is done by randomly dropping a fraction $f$ of the shower segments and then scaling up the response in each cell by $1/f$. To focus solely on the calorimeter response, all upstream material, such as the inner tracker and solenoid iron, was removed, and the magnetic field was turned off, allowing showers from charged particles to propagate radially.

In this study, we apply superresolution exclusively in the pseudorapidity ($\eta$)-azimuthal angle ($\phi$) plane, keeping the calorimeter layers unchanged. We use two datasets: the first is a simplified set where a single electron is fired with $p_T \in [50, 51]\ \mathrm{GeV}$, $\eta \in [-0.01, 0.01]$, and $\phi \in [-\pi, \pi]$. The second dataset involves multiple particles with angular separation on the scale of the calorimeter cells. The particles are arranged around a principal axis chosen uniformly in $\eta \in [-2.5, 2.5]$ and $\phi \in [-\pi, \pi]$. In three out of four cases, an electron is fired along the principal axis with $p_T \in [20, 50]\ \mathrm{GeV}$. In addition, $N$ photons are fired with $p_T \in [5, 25]\ \mathrm{GeV}$, where $N$ is chosen randomly from 0 to 3 (and required to be nonzero in case of no electron). Each photon is displaced from the principal axis by $\pm \mathrm{Gaus}(4\cdot\delta, \delta)$, where the sign is chosen randomly and $\delta$ is the angular width of a cell in the high-resolution ECAL1. For the single-electron dataset, the cell resolution is downsampled by factors, $f_\eta = f_\phi = 2$, and for the multiparticle dataset, by $f_\eta = f_\phi = 4$. Table~\ref{tab:CaloLayers} details the noise levels and granularity used for each calorimeter layer in this study.

\begin{table}[b]
    \caption{Noise levels and granularity of the three electromagnetic calorimeter layers in both low and high resolution. The noise is only included for the low-granularity cells.}
    \label{tab:CaloLayers}
    \begin{ruledtabular}
    \begin{tabular}{llccc}
    Layer & & ECAL1 & ECAL2 & ECAL3 \\
    \hline
    $\sigma_\mathrm{noise}$ [MeV] & (Low) & 13 & 34 & 41 \\
    $\eta, \phi$ segmentation & (Low) & 128 & 128 & 64 \\
    single electron & (High) & 256 & 256 & 128 \\
    $\eta, \phi$ segmentation & (Low) & 64 & 64 & 32 \\
    multiparticle & (High) & 256 & 256 & 128 \\
    \end{tabular}
    \end{ruledtabular}
\end{table}

\begin{figure*}[t]
  \centering
  \hfill
  \includegraphics[width=0.5\textwidth]{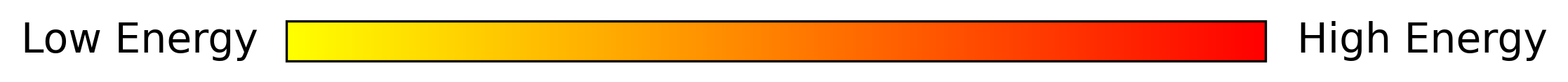}%
  \hspace*{0.13\textwidth}

  \vspace{0.5em}
  \begin{subfigure}{.25\textwidth}
    \centering
    \includegraphics[width=\linewidth]{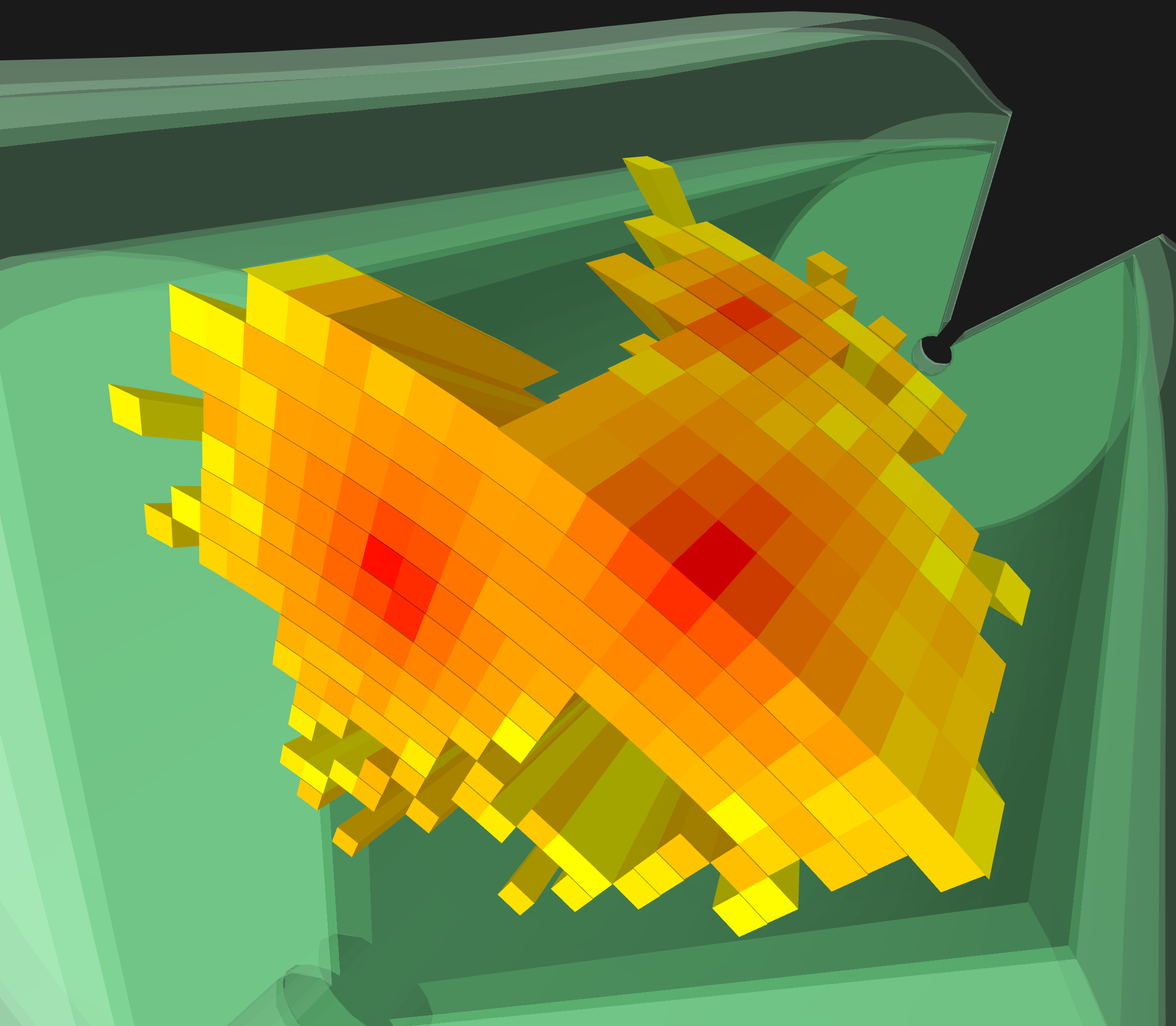}
    \caption{}
  \end{subfigure}%
  \hspace{5pt}
  \begin{subfigure}{.25\textwidth}
    \centering
    \includegraphics[width=\linewidth]{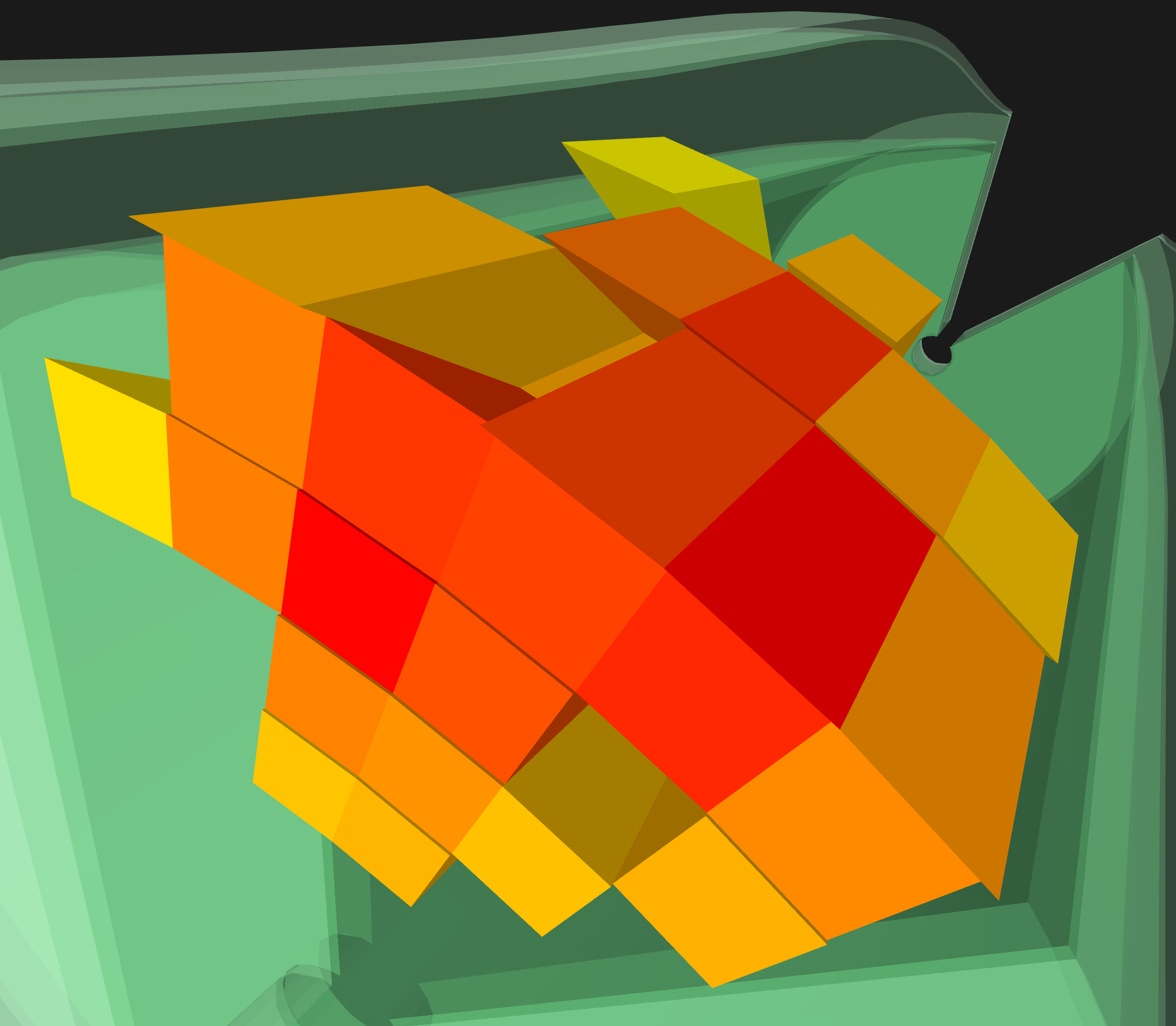}
    \caption{}
  \end{subfigure}%
  \hspace{5pt}
  \begin{subfigure}{.25\textwidth}
    \centering
    \includegraphics[width=\linewidth]{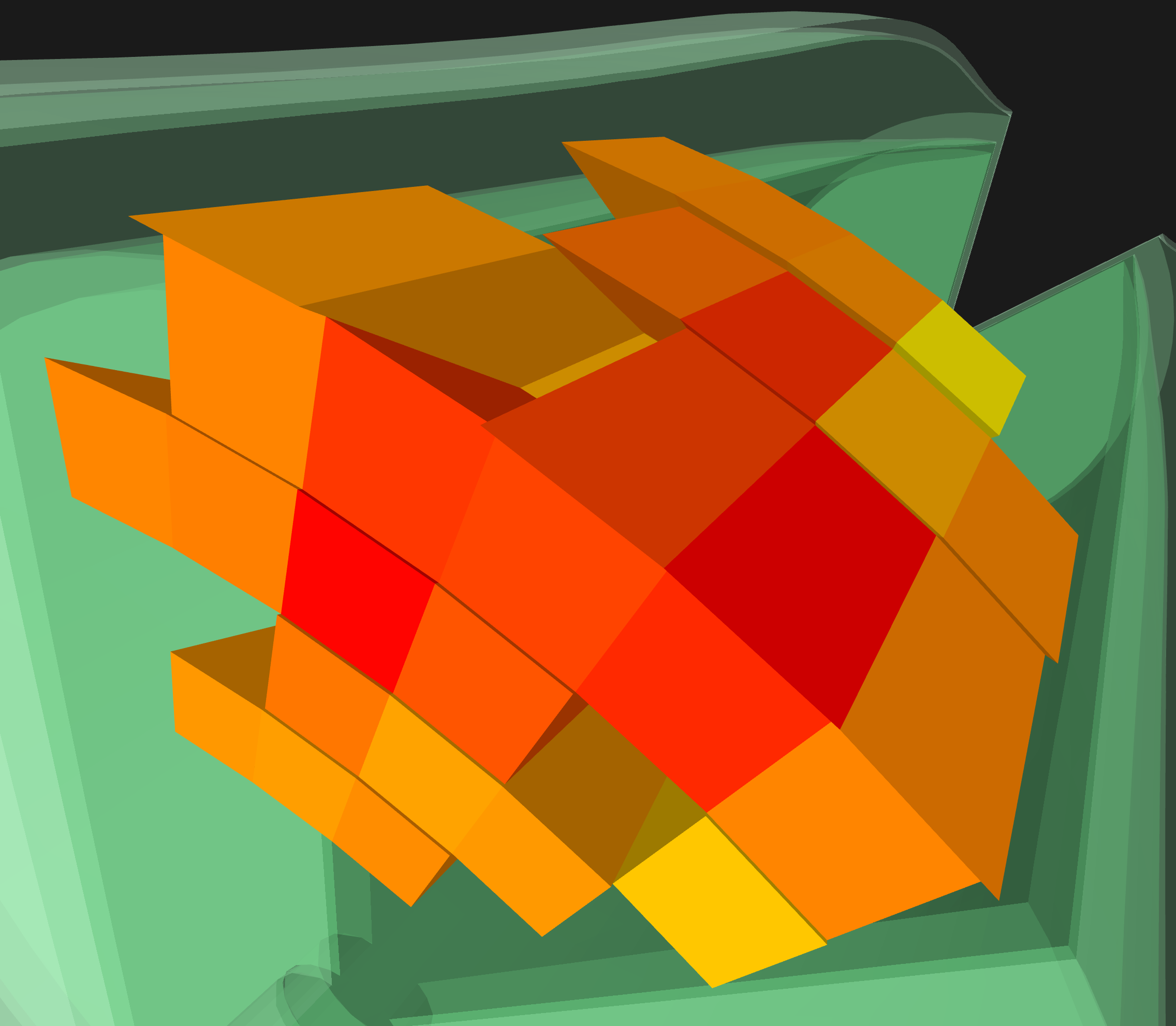}
    \caption{}
  \end{subfigure}
  \caption{Phoenix event display~\cite{phoenix} of one electron and two photons in \textsc{cocoa}, demonstrating the dataset creation procedure. Here we see (a) the truth high-resolution configuration, (b) the truth low-resolution configuration, and (c) the measured low-resolution configuration which also contains noise.}
  \label{fig:lr_hr_datagen}
\end{figure*}

The particles interact with the high-resolution detector, initially with zero noise. The high-resolution cells are then combined to form low-resolution cells, to which noise is applied to simulate electronic noise. This procedure is illustrated in Fig.~\ref{fig:lr_hr_datagen}. Subsequently, the built-in topological clustering algorithm in \textsc{cocoa}, which resembles that of ATLAS, is used to cluster calorimeter hits based on energy-over-noise ratio thresholds. Only clustered cells with positive energy values are retained. Since our focus is on electromagnetic showers from electrons and photons, we limit our study to the ECAL layers.

\section{Methods} \label{sec:methods}
\subsection{Continuous normalizing flows}
Continuous normalizing flow~\cite{NEURIPS2018_69386f6b} is a type of generative model that transforms a simple base probability distribution $p_0 (x_0)$ into a more complex target distribution $p_1 (x_1)$ through a continuous sequence of invertible transformations. This transformation is governed by the following neural ordinary differential equation:
\begin{equation}
    \frac{dx}{dt} = v_{\theta} \left( x, t \right),
\end{equation}
where $v_\theta$ is the vector field parametrized by a neural network and $t \in \left[ 0, 1 \right]$. Flow matching, as proposed in~\cite{lipman2023flow}, is a technique for training continuous normalizing flows that aims to match the learned flow $v_{\theta} \left( t, x\right)$ induced by the differential equations of the model with a target flow $u_t\left( x \right)$. The flow-matching objective can be defined as follows:
\begin{equation}
    L_{FM} \left( \theta \right) = E_{t, p_t(x)} \lVert v_{\theta} \left( t, x\right) - u_t \left( x \right) \rVert ^2 .
\end{equation}
The objective, however, is intractable because the target flow $u_t$, that generates the desired probability paths between $p_0$ and $p_1$, is typically not available in closed form. As a solution,~\cite{lipman2023flow} also proposed a conditional flow-matching objective where $u_t$ and $p_t$ can be constructed in a sample-conditional manner:
\begin{equation}
    L_{CFM} \left( \theta \right) = E_{t, q(z), p_t(x|z)} \lVert v_{\theta} \left( t, x\right) - u_t \left( x | z\right) \rVert ^2,
\end{equation}
where $t \in \left[ 0, 1 \right]$, $z \sim q\left(z\right)$, and $x \sim p_t\left(x|z\right)$. $L_{CFM}$ is fully tractable and the authors showed that optimizing it is equivalent to optimizing the $L_{FM}$ objective. The probability path $p_t$ and the target flow $u_t$ can be defined in various ways~\cite{lipman2023flow, tong2024improving}. For our studies, we use the formulation from the original paper~\cite{lipman2023flow}:
\begin{equation}
\begin{aligned}
    p_t \left( x | z \right) &= \mathcal{N} \left( x | t x_1, \left( t \sigma - t + 1 \right)^2 \right), \\
    u_t \left( x | z \right) &= \frac{x_1 - \left( 1 - \sigma \right)x}{1 - \left( 1 - \sigma \right)t} .
\end{aligned}
\end{equation}
This defines a straight path between the standard normal and a Gaussian distribution of width $\sigma$ around $x_1$. We chose $\sigma$ to be $10^{-4}$ following the original paper. We used the \textsc{torchdiffeq}~\cite{torchdiffeq} library with the Dormand-Prince (dopri5) solver for generation.

\subsection{Superresolution model}

\begin{figure}[t]
    \centering
    \includegraphics[width=\linewidth]{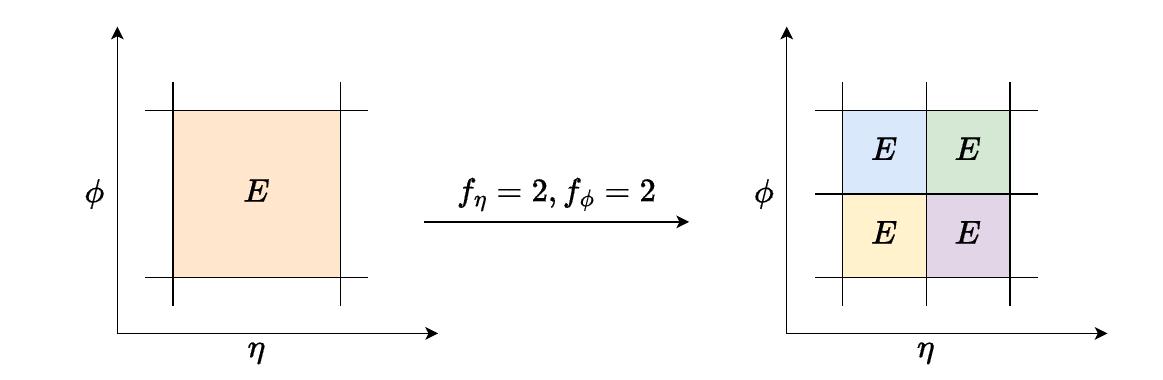}
    \caption{Creation of high-resolution cell in our approach. Each low-resolution cell is split in $\eta$ and $\phi$ by the superresolution factor $f_\eta$ and $f_\phi$, respectively. They get their new angular coordinates and the parent low-resolution energy $\left( E \right)$ is copied to serve as the conditional inputs. Here, we see the procedure for $f_\eta = 2$ and $f_\phi = 2$. Each cell is a node in the graph, but, for illustrative purpose, we are denoting them as images.}
    \label{fig:lr_hr_cell_treatment}
\end{figure}

The core concept of our superresolution model is inspired by the work~\cite{9887996}. To allow for irregular cell geometry and to cope with the sparsity of the recorded energy deposits, we represent the data as graphs. Each low-resolution graph is expanded into a high-resolution graph, where each low-resolution cell is divided into $f_{\eta} \times f_{\phi}$ high-resolution cells. For each high-resolution cell, the target is its truth energy value, while the corresponding low-resolution energy serves as the conditional input, as illustrated in Fig.~\ref{fig:lr_hr_cell_treatment}. Although nearest-neighbor edges among nodes in the high-resolution graph would, in general, be motivated for larger graphs, we model them as fully connected for simplicity.

\begin{figure*}[t]
    \centering
    \includegraphics[width=0.6\textwidth]{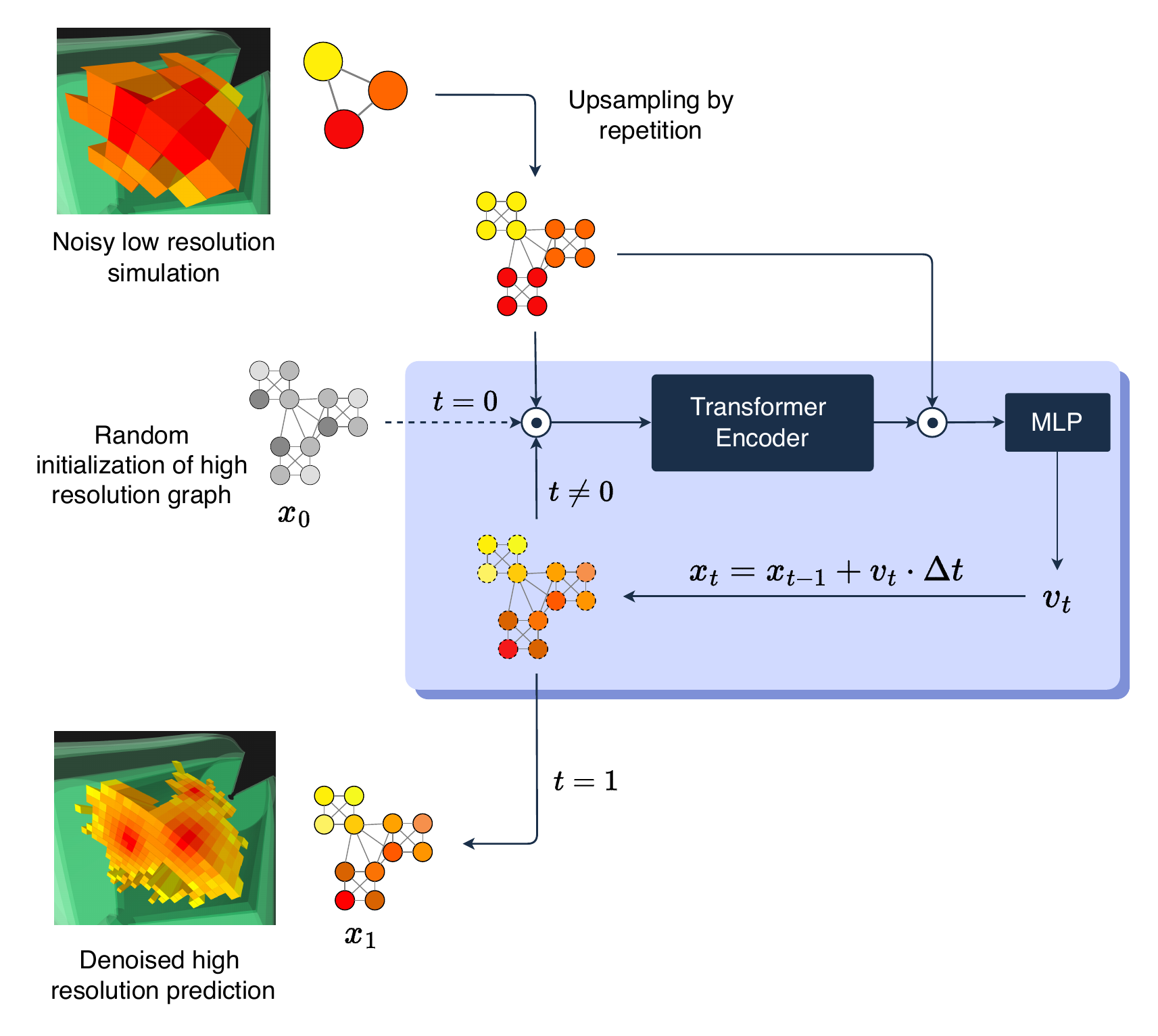}
    \caption{The superresolution architecture. $\odot$ indicates concatenation.}
    \label{fig:sup_res_model}
\end{figure*}

Our model employs a transformer-based architecture to predict the vector field $v_t$, as illustrated in Fig.~\ref{fig:sup_res_model}. The process begins by upsampling the low-resolution graph through repetition, which serves as the conditional input. Subsequently, a copy of the high-resolution graph, initialized with random values $x_0$, is created. These two high-resolution graphs are then stacked, providing each node with both the upsampled information and the randomly initialized values. This combined input is fed into a transformer network, generating a neighborhood-aware representation that aids in understanding the energy deposition profile of the incident particles. Following this, a multilayer perceptron (MLP) predicts the vector field $v_t$ for each node. To advance the graph, we integrate $v_t$ over a small time interval $\Delta t$, yielding $x_{0 + \Delta t}$. This updated graph $x_{0+\Delta t}$ is then used as input, along with the conditional information, for the subsequent transformer iteration, and the process of predicting $v_t$ is repeated. Finally, at $t=1$, the output $x_1$ represents the denoised, superresolved graph.

The hyperparameters used during training are summarized in Table~\ref{tab:sr_hyperparameters}. For the single-electron study, the training, validation, and test datasets consisted of 1\,000\,000, 3000, and 10\,000 events, respectively. In the multiparticle study, we utilized 250\,000 events for training, 3000 for validation, and 10\,000 for testing.

\begin{table}[t]
    \caption{Hyperparameters used in the superresolution model.}
    \label{tab:sr_hyperparameters}
    \begin{ruledtabular}
    \begin{tabular}{lr}
        Hyperparameter & \\
        \hline
        Batch size & 25 \\
        Optimizer & AdamW~\cite{Loshchilov2017DecoupledWD} \\
        Learning rate & 0.001 \\
        Number of epochs & 100 \\[0.8em]
        Number of time steps & 25 \\
        Number of transformer layers & 4 \\
        Number of trainable parameters & 4.2M \\
    \end{tabular}
    \end{ruledtabular}
\end{table}

\subsection{Ensemble sampling}
\label{sec:ensemble_sampling}
Generation with continuous normalizing flow is inherently a stochastic process. While this randomness is often beneficial in applications like image generation, where variation is desired, it poses a challenge in the superresolution of calorimeter data, where there is a specific deterministic target. Ideally, the network's output should fluctuate around the true target if functioning correctly. To address this, we employ an ensemble approach. Unlike typical ensemble methods that involve training multiple copies of the model, our approach simply involves passing the same conditional input through the same model multiple times to produce the ensemble output, which is then averaged to obtain the final prediction. This technique helps reduce the impact of fluctuations and brings the output closer to the true target. However, it is important to acknowledge that due to the stochastic noise inherently added by the calorimeter, recovering the exact true energy is not possible; we can only approximate it.

\subsection{Particle-flow model}
Machine learning-based particle-flow models for LHC-like environments have been extensively explored~\cite{Kieseler_2020, Qasim:2019otl, qasim2021multi, qasim2022end, pata2021mlpf, Pata:2022wam, Mokhtar:2023fzl, Pata:2023rhh, pflow, hgpflow}. Among these, \textit{HGPflow}~\cite{pflow, hgpflow} distinguishes itself by explicitly learning the energy assignment matrix between inputs and outputs rather than treating the reconstruction as a black-box regression. This approach offers a high degree of interpretability, as it allows for the direct inspection of how recorded energy in the detector is assigned to reconstructed particles. To better understand the impact of superresolution, we aimed for a similarly interpretable model. Here, we present a novel particle-flow architecture that builds on the principles of HGPflow but replaces the \textit{iterative learning} approach with a \textit{supervised attention} mechanism.

PFlow algorithms usually combine track and calorimeter inputs. However, since we are primarily interested in studying the impact of calorimeter SR, we do not include tracks in the input. To indicate this, we refer to our model as PFlow\textsuperscript{--}, since the architecture itself can be used in the same capacity as HGPflow.

\begin{figure*}[t]
    \centering
    \includegraphics[width=0.7\textwidth]{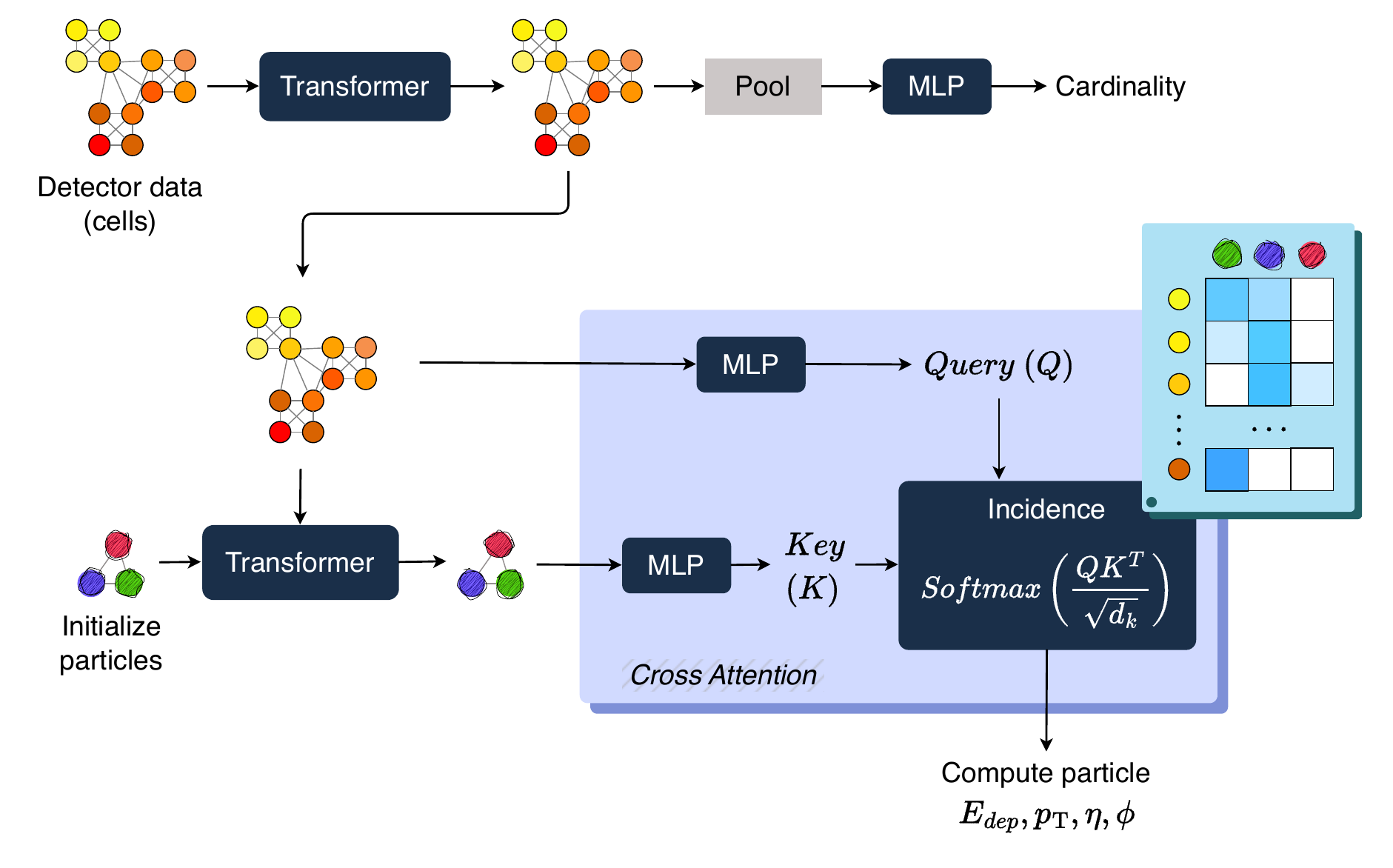}
    \caption{The PFlow\textsuperscript{--} architecture. The core concept is similar to~\cite{pflow}, but instead of learning the incidence matrix through iterative refinements, we predict it as a standard cross attention.}
    \label{fig:pflow_model}
\end{figure*}

Figure~\ref{fig:pflow_model} illustrates the high-level view of the architecture. The model processes detector data from cells using a transformer network to generate a neighborhood-aware representation. The transformer's output is processed by two branches. One branch predicts the particle cardinality using an MLP, which treats this prediction as a classification task rather than regression. While predicting cardinality as a global metric is not the most robust approach, and many existing models~\cite{pflow, pata2021mlpf} handle this as a padding problem (predicting a flag for each particle to indicate if it's real or padded), we opted for a simpler classification approach given our lower cardinality and simpler environment.

The second branch focuses on kinematic prediction. It achieves this by learning the fractional energy distribution of each cell over the particles as shown in Eq.~(\ref{eq:incidence}), where $E_{ai}$ represents the energy deposited by particle $a$ in cell $i$. This distribution, referred to as the incidence matrix in HGPflow~\cite{pflow}, was originally defined for topological clusters rather than individual cells. By using cells directly, we retain more detailed information, as topological clusters aggregate the features of multiple cells, potentially losing fine-grained details. Working with cells increases the complexity, since the cardinality can be an order of magnitude or more above that of clusters. In exchange it eliminates the dependency on topoclustering, which is not guaranteed to be an optimal algorithm. We implement this approach using an attentionlike framework, where particles and cells are processed through two MLPs to obtain the key ($K$) and query ($Q$), respectively. The resulting attention weights are computed as usual [Eq.~(\ref{eq:attention})] and supervised to match the target incidence matrix. Given that we operate on sets, permutation invariance of particles is essential, and we achieve this using Hungarian matching~\cite{kuhn1955hungarian}:
\begin{equation}
\label{eq:incidence}
\begin{aligned}
\text{target incidence},\quad I_{ai} &=
\frac{
  \begin{array}{c}
  \text{energy deposited by particle } a \\[2pt]
  \text{in cell } i
  \end{array}
}{
  \begin{array}{c}
  \text{total energy deposited} \\[2pt]
  \text{in cell } i
  \end{array}
} \\
&= \frac{E_{ai}}{\sum_a E_{ai}} ,
\end{aligned}
\end{equation}
\begin{equation}
\label{eq:attention}
\begin{aligned}
\text{predicted incidence} &= \text{attention weights} \\
&= \mathrm{Softmax} \left( \frac{QK^T}{\sqrt{d_K}} \right) .
\end{aligned}
\end{equation}

Once the energy-based incidence matrix is determined, the \textit{proxy} kinematics of the reconstructed particles can be calculated from the associated cells using the approach outlined in~\cite{pflow}:
\begin{equation}
\begin{aligned}
    \label{eq:ptetaphi_computation}
    &\hat{E}_a = \sum_i \left( I_{ai} \cdot E_i \right); \qquad
    \hat{p}_{\mathrm{T}_a} = \frac{\hat{E}_a}{\cosh \hat{\eta}_a}\, (\text{zero mass}), \\
    &\hat{\eta}_a = \frac{1}{\hat{E}_a}\sum_i \left( I_{ai} \cdot E_i \cdot \eta_i \right); \qquad
    \hat{\phi}_a = \frac{1}{\hat{E}_a}\sum_i \left( I_{ai} \cdot E_i \cdot \phi_i \right).
\end{aligned}
\end{equation}

In HGPflow~\cite{pflow}, these proxy properties are further refined using MLPs, a step we omit in our default configuration. Keeping the kinematics an explicit function of the incidence matrix and the cell coordinates ensures that any difference in performance between the two resolutions can be attributed to the input cells rather than to a learned correction. This choice makes the comparison transparent, but it also ties the angular resolution to the granularity of the input, since the predicted position can only be expressed as an energy weighted combination of cell centers. To quantify the effect of this restriction, we additionally train a variant in which the proxy kinematics are refined by a correction network, following~\cite{pflow}. The correction is implemented as a residual on top of Eq.~(\ref{eq:ptetaphi_computation}). It takes the full per-particle representation as input, and is initialized to zero, so that training begins from the uncorrected estimator. We apply it to both the low- and the high-resolution models and report the results.

We trained two identical models using the same set of hyperparameters: one with the low-resolution measured cells and the other with the high-resolution predicted cells. The particle-flow models were trained on the superresolution predictions for the 250\,000 training events, and all results are reported on a separate sample of 10\,000 events that was not used to train either the superresolution model or the particle-flow models. For the incidence loss, we followed a procedure similar to that in~\cite{pflow}, pairing truth and predicted particles using the Hungarian algorithm with the Kullback-Leibler divergence between their incidence values as the metric. Cardinality loss was handled using cross-entropy, and the sum of the incidence and cardinality losses was used for optimization. The same procedure was used for the variants with the correction network, with the addition of per-particle kinematic terms for the angular coordinates and the energy, without which the correction network receives no gradient. Their weights were chosen so that each term contributes comparably at the start of training, and these models were initialized from the corresponding uncorrected model and trained for a further 60 epochs. The hyperparameters used for training are summarized in Table~\ref{tab:pf_hyperparameters}.

\begin{table}[t]
    \caption{Hyperparameters used in the particle-flow model.}
    \label{tab:pf_hyperparameters}
    \begin{ruledtabular}
    \begin{tabular}{lr}
        Hyperparameter & \\
        \hline
        Batch size & 30 \\
        Optimizer & AdamW~\cite{Loshchilov2017DecoupledWD} \\
        Learning rate & 0.001 \\
        Number of epochs & 100 \\[0.8em]
        Number of transformer encoder layers & 3 \\
        Number of transformer decoder layers & 4 \\
        Number of trainable parameters & 333K (+19K) \\
    \end{tabular}
    \end{ruledtabular}
\end{table}

\section{Results} \label{sec:results}
\subsection{Single electron}

\begin{figure*}[t]
  \centering
  \includegraphics[width=0.9\textwidth]{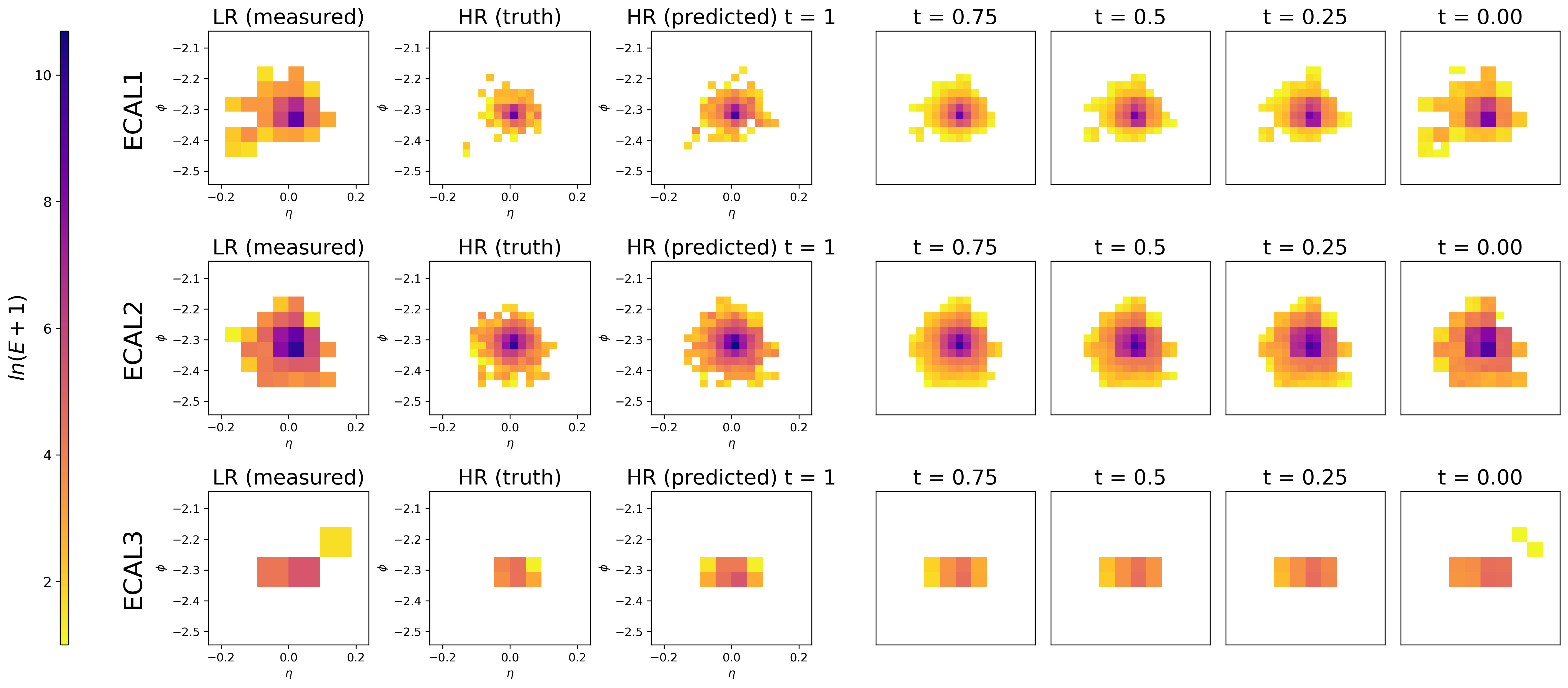}
  \caption{Superresolution applied to a single-electron event from the test set. The three rows depict the superresolution process for the three ECAL layers. For each layer, the measured LR cell energy distribution, the target HR cell energy distribution, and the predicted HR are shown. The predicted HR cells closely match the truth HR ones. Additionally, the time evolution of the cell energy distribution is also illustrated.}
  \label{fig:se_evolution}
\end{figure*}

Figure~\ref{fig:se_evolution} provides a qualitative view of superresolution applied to a single-electron event. The predicted high-resolution cell energies closely align with the true values, and the time evolution of these predicted energies is also shown. Given that the target definition involves a correction to the input energy, the $t=0$ state is not entirely random but instead closely reflects the conditional input. The network also manages to suppress noise as seen from the figure.

\begin{figure*}[t]
  \centering
  \includegraphics[width=0.85\textwidth]{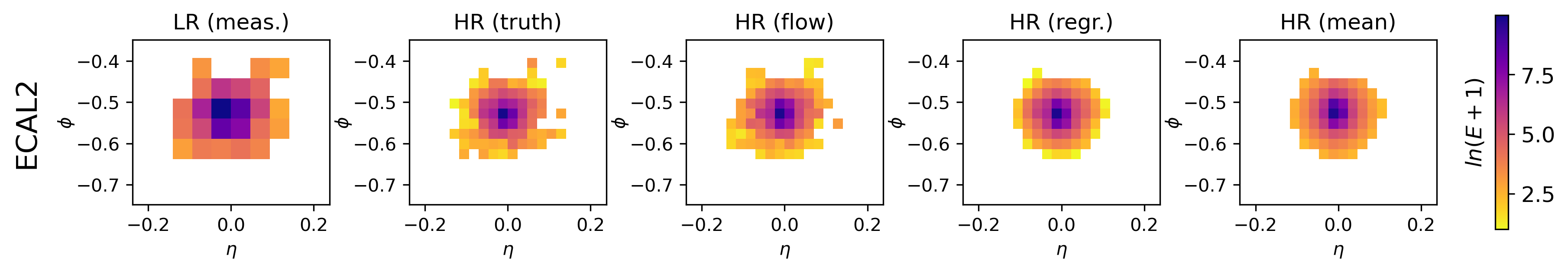}
  \caption{Comparison of flow and regression model outputs for superresolution on a single-electron event from the test set.}
  \label{fig:flow_vs_regression}
\end{figure*}

To compare our generative architecture with a deterministic one, we trained a variant with a direct regression objective rather than a flow-matching objective. The regression model has an identical setup except that the time step and current state embedding inputs are dropped. Figure~\ref{fig:flow_vs_regression} shows qualitatively how the regression model's output compares to the flow model output. While the flow model attempts to generate the local fluctuations in energy, the regression model produces a smooth energy distribution. For reference, we include an evaluation of the shower mean computed over all truth HR examples, which behaves similarly to the regression model output.

\begin{figure*}[t]
\centering
\begin{subfigure}{.33\textwidth}
  \centering
  \includegraphics[width=\linewidth]{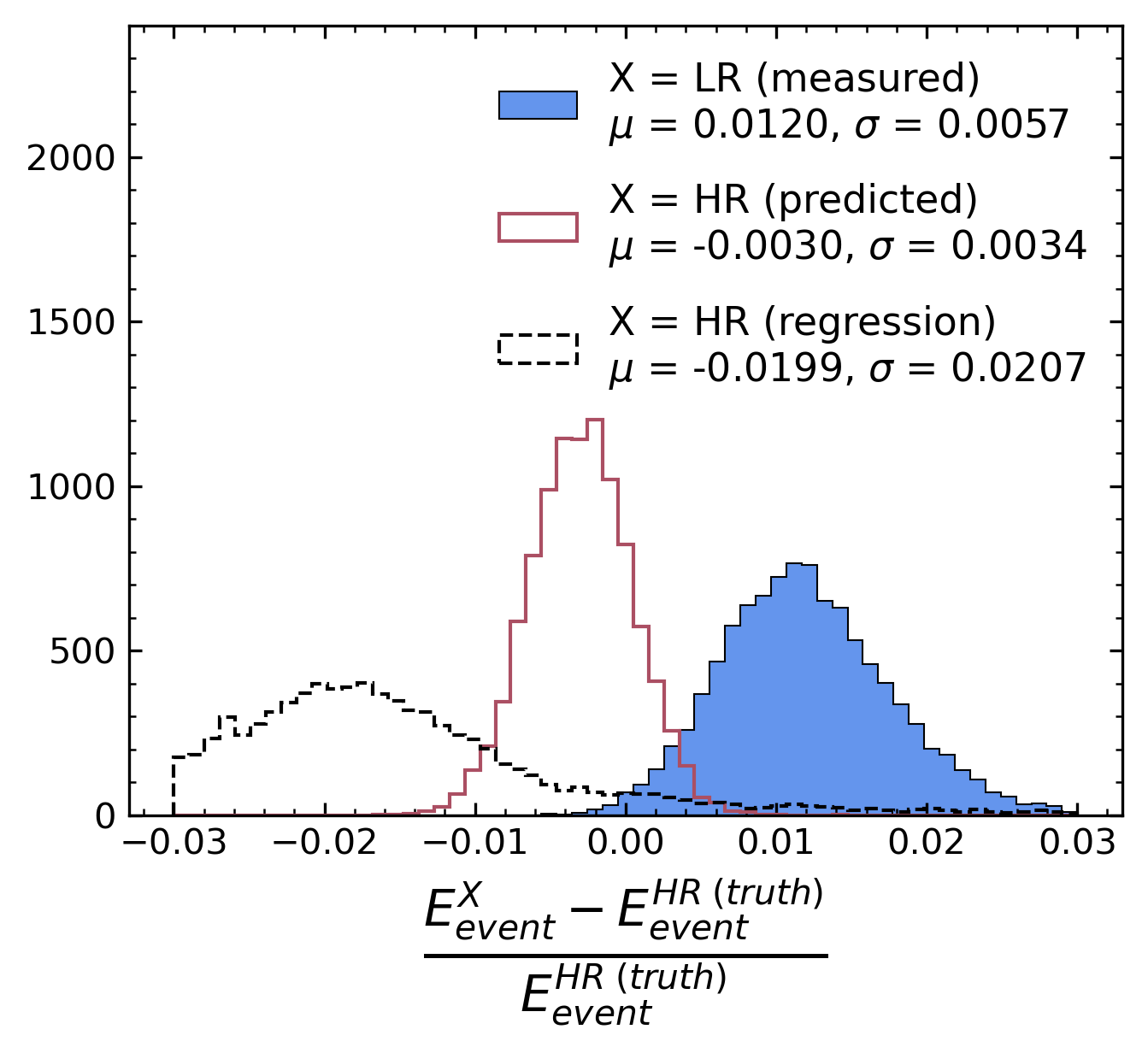}
  \caption{}
  \label{fig:se_residual}
\end{subfigure}%
\hspace{5pt}
\begin{subfigure}{.40\textwidth}
  \centering
  \raisebox{1.7em}{%
    \includegraphics[width=\linewidth]{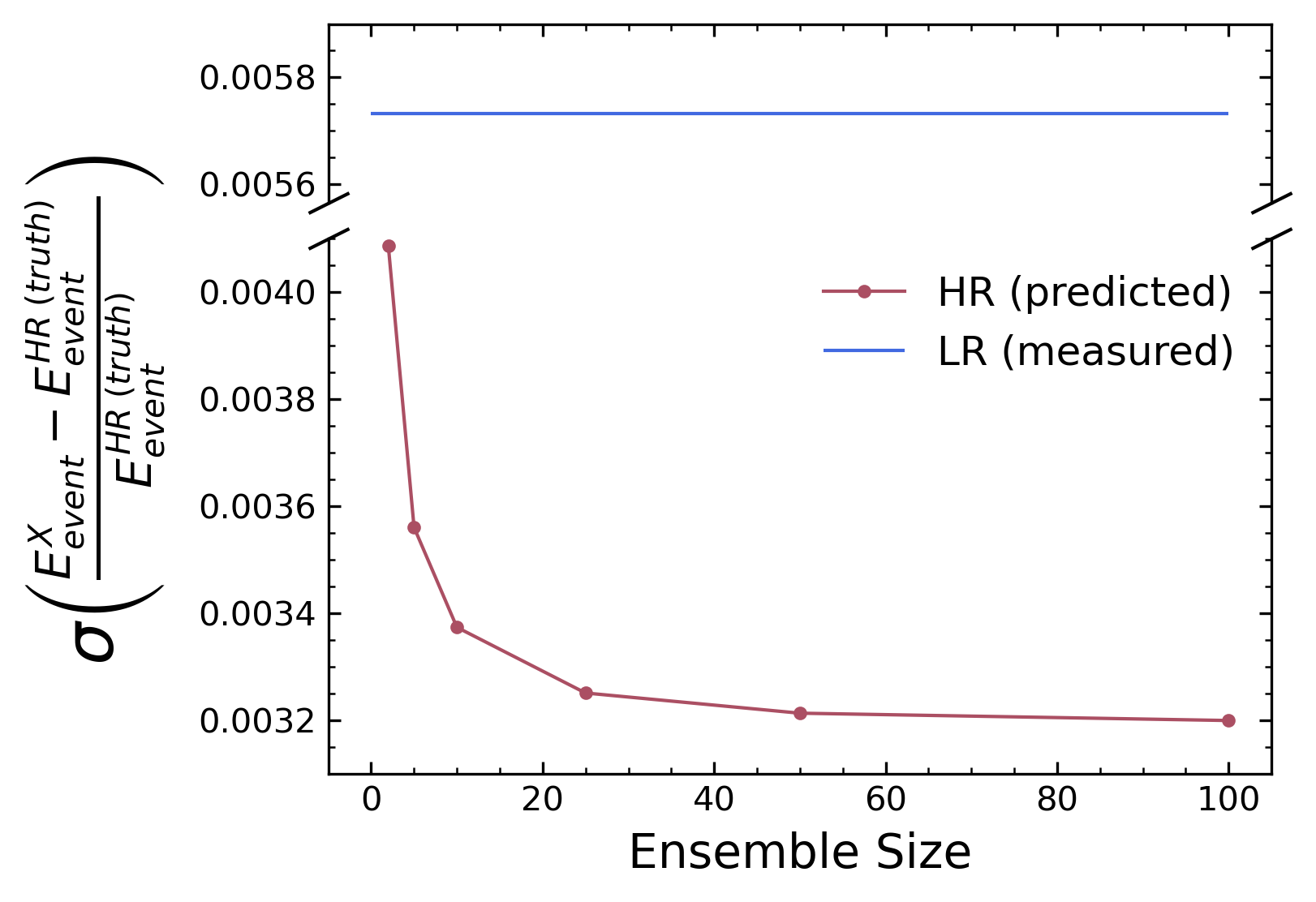}%
  }
  \caption{}
  \label{fig:se_ensemble_size}
\end{subfigure}
\caption{(a) The event-level relative energy residual for the LR measured cell energy and the HR predicted cell energies with an ensemble size of 10. (b) The one-$\sigma$ width of the event-level residual distributions for the predicted cell energies as a function of the ensemble size.}
\label{fig:se_res_ensemble}
\end{figure*}

To quantitatively assess the impact on reconstruction, we estimate the electron's energy by summing the energies of all cells in the event. We examine the energy residuals using both the measured low-resolution cell energies and the predicted high-resolution cell energies, with the target being the sum of the true high-resolution cell energies. Figure~\ref{fig:se_residual} illustrates these residual distributions. As discussed in Sec.~\ref{sec:ensemble_sampling}, the predicted high-resolution energies result from ensemble sampling, with an ensemble size of 10 in this instance. The figure shows that the high-resolution residual is much closer to zero and exhibits a narrower spread. The peak shift in the low-resolution distribution is due to the asymmetric noise contribution, as we retain only the cells with positive energy. The network effectively adapts to this asymmetric noise, demonstrating an improvement in energy resolution. The performance of the regression model is also shown by the dashed line in Fig.~\ref{fig:se_residual}. Its tendency to predict the marginal distribution is reflected in the wider residual distribution. The negative shift reflects its attempt at removing positively biased noise without the capability of producing local fluctuations.

We also examine how the performance depends on the ensemble size. As expected, the position of the residual distribution remains unchanged regardless of the ensemble size, and the width of the distribution varies. Figure~\ref{fig:se_ensemble_size} illustrates how the 1$\sigma$ width of the residual distribution decreases as the ensemble size increases, with the improvement saturating above an ensemble size of 20. For all our studies, both for the single-electron sample and the multiparticle sample, we opted for an ensemble size of 10, to keep the computational costs lower while still benefiting from the advantages of ensemble sampling.

\begin{figure*}[t]
    \centering
    \includegraphics[width=0.85\textwidth]{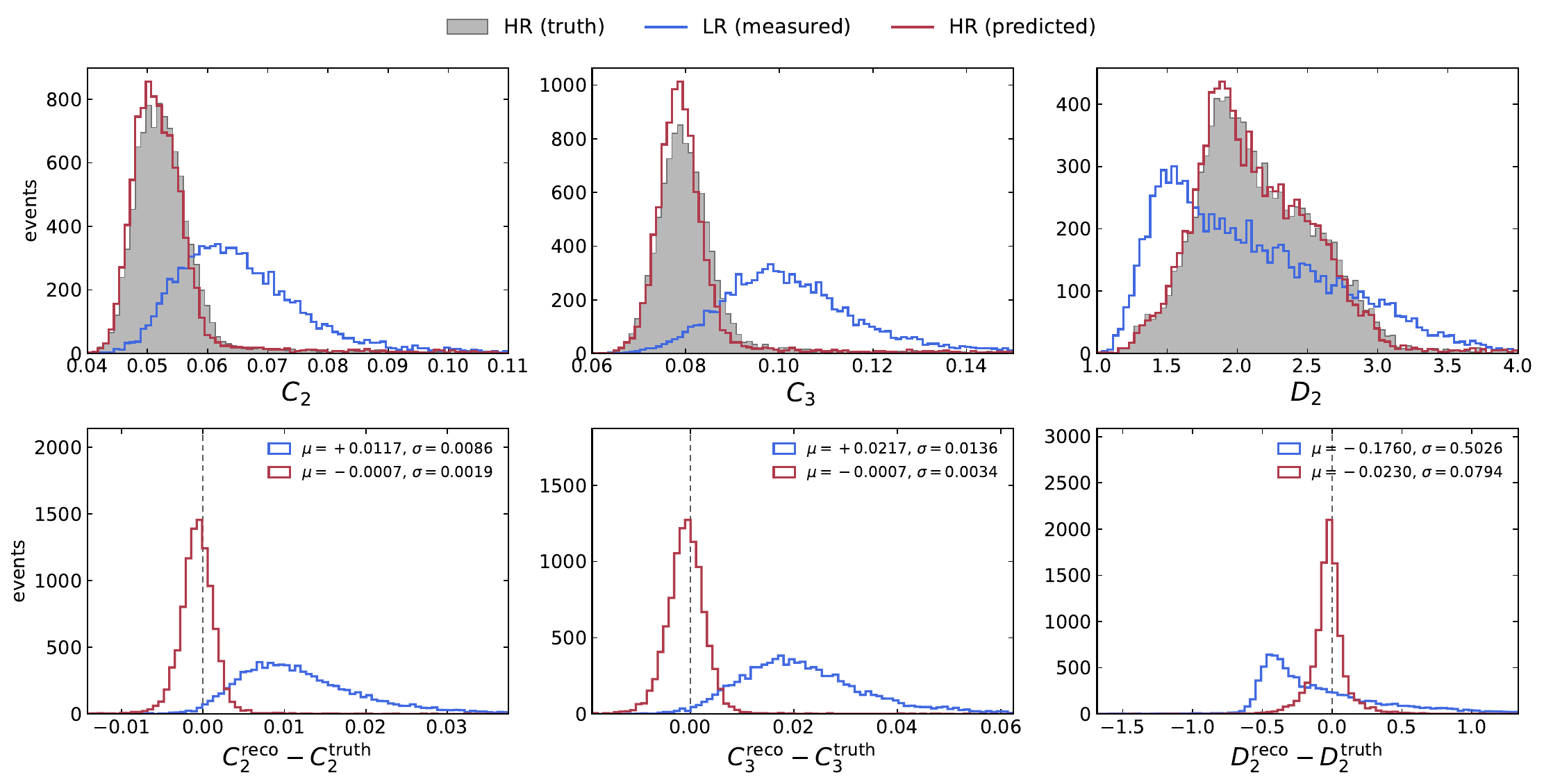}
    \caption{Substructure distributions for the truth high-resolution, measured low-resolution and predicted high-resolution cell energies (upper row), and the corresponding per-event residuals with respect to the truth high-resolution value (lower row). All sets of cells are computed on the same events. Here $\mu$ and $\sigma$ denote the median and the interquartile range divided by $1.349$.}
    \label{fig:se_substructure}
\end{figure*}

In addition, we analyzed the energy correlator observables $C_2$, $C_3$, and $D_2$~\cite{Larkoski:2013eya,Larkoski:2014gra} at the cell level for both high-resolution and low-resolution data. Unlike the total energy, which is a sum over cells and therefore unchanged by downsampling, these observables depend on the spatial distribution of the energy and are sensitive to granularity. The upper row of Fig.~\ref{fig:se_substructure} presents the resulting distributions, with lower values indicating more substructure. The predicted high-resolution cells closely match the true distribution and offer significant improvements over the noisy low-resolution data across all three substructure observables considered. The lower row shows the per-event residuals, which for the predicted high-resolution cells are narrower than the spread of the observables themselves, so the prediction follows the individual event rather than reproducing the ensemble distribution.

\subsection{Multiple particles}

For the multiparticle sample, the most exciting aspect of the study is the impact on PFlow performance. We begin by examining the cardinality performance of the two networks, as summarized in Table~\ref{tab:pf_cardinality}. Across the board, the network trained with predicted high-resolution cells shows either comparable or improved performance relative to the one trained with low-resolution measured cells. Both networks almost perfectly identify single-particle events, but as the true cardinality increases, their performance degrades, though at a slower rate for the network trained with high-resolution cells. When the true cardinality reaches its maximum, both networks exhibit similar performance, as this represents an edge-case scenario.

\begin{table}[t]
    \caption{Cardinality prediction with the network trained on the LR cells with measured energy, compared with the network trained on the HR cells with predicted energy.}
    \label{tab:pf_cardinality}
    \begin{ruledtabular}
    \begin{tabular}{ccc}
        & \multicolumn{2}{c}{Accuracy} \\
        \cline{2-3}
        Truth cardinality & LR (measured) (\%) & HR (predicted) (\%) \\
        \hline
        1 & 99.38 & \textbf{99.41} \\
        2 & 88.52 & \textbf{91.02} \\
        3 & 77.62 & \textbf{81.97} \\
        4 & 81.32 & \textbf{81.42} \\
    \end{tabular}
    \end{ruledtabular}
\end{table}

\begin{figure*}[t]
\centering
\begin{subfigure}{0.36\textwidth}
  \centering
  \raisebox{0.48em}{%
    \includegraphics[width=0.95\linewidth]{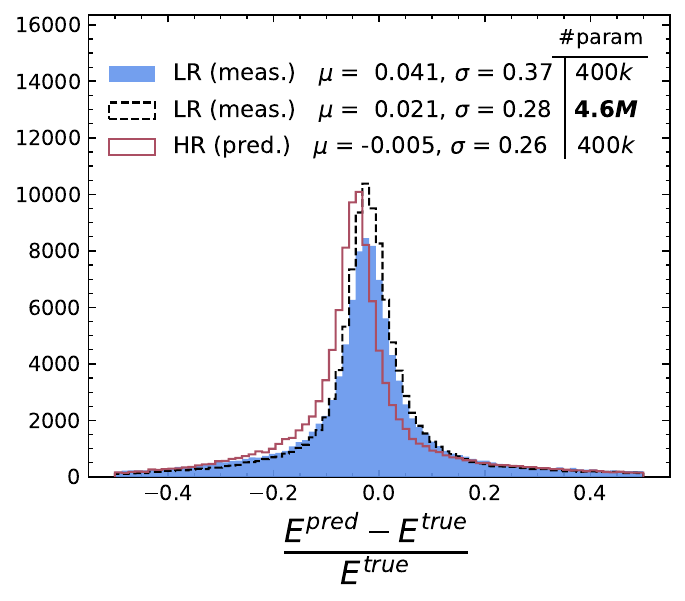}%
  }
  \caption{}
  \label{fig:pf_res_e}
\end{subfigure}%
\hspace{5pt}
\begin{subfigure}{0.36\textwidth}
  \centering
  \includegraphics[width=0.95\linewidth]{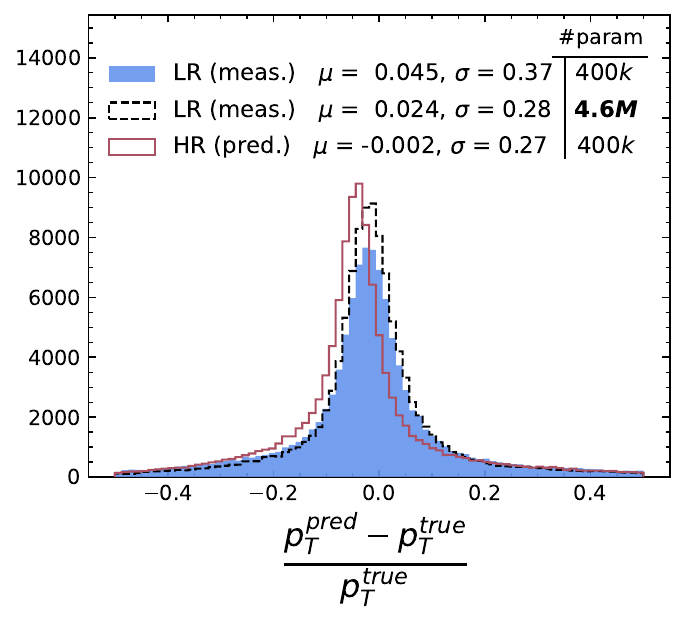}
  \caption{}
  \label{fig:pf_res_pt}
\end{subfigure}

\begin{subfigure}{.36\textwidth}
  \centering
  \includegraphics[width=0.95\linewidth]{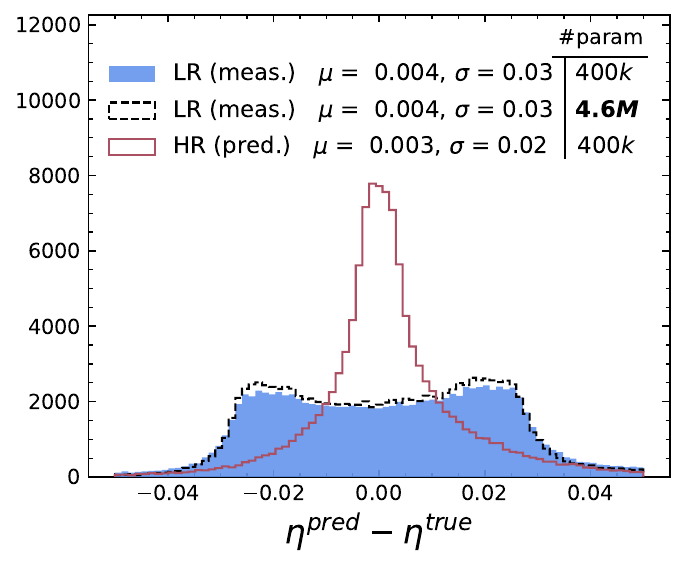}
  \caption{}
  \label{fig:pf_res_eta}
\end{subfigure}%
\hspace{5pt}
\begin{subfigure}{.36\textwidth}
  \centering
  \includegraphics[width=0.95\linewidth]{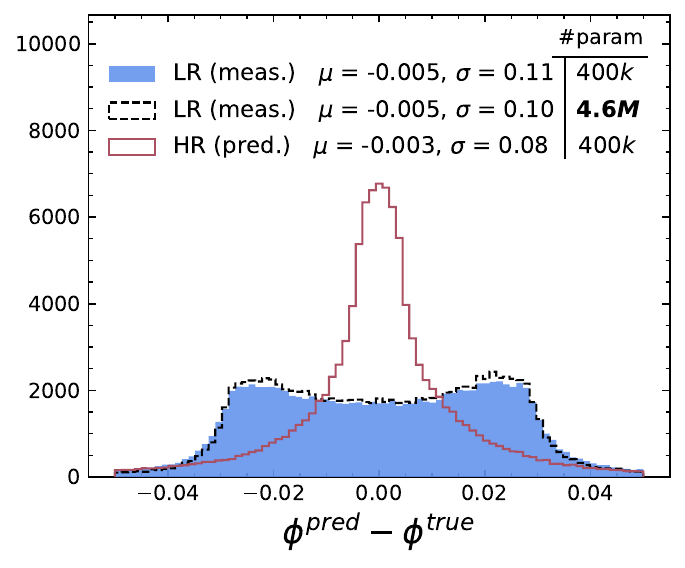}
  \caption{}
  \label{fig:pf_res_phi}
\end{subfigure}
\caption{Particle-flow performance. (a) The relative energy residuals, (b) the relative transverse momentum residuals, (c) the pseudorapidity residuals, and (d) the azimuthal angle residuals between the truth particles and the reconstructed particles with PFlow model trained on measured low-resolution cells and the predicted high-resolution cells. The dashed line indicates the performance of an additional PFlow model trained on the low-resolution cells but with over 10$\times$ more parameters.}
\label{fig:pf_res}
\end{figure*}

Next, we evaluate the performance of the computed particle $p_T$, $\eta$, and $\phi$ values derived from the incidence predictions. Ideally, the kinematic model would receive the predicted cardinality as input during inference. However, to isolate the performance of the kinematics predictor from the cardinality predictor, we instead use the true cardinality as input. To evaluate the kinematics at the particle level, we pair the true and predicted particles per event through Hungarian matching, using the Kullback-Leibler divergence loss between their incidence values as the cost matrix. The $p_T$, $\eta$, and $\phi$ residuals after matching are shown in Fig.~\ref{fig:pf_res}. Both models perform well in modeling the kinematic values, but the model trained on predicted high-resolution cells shows a marked improvement in angular resolution.

While both PFlow models have 400k parameters, one may argue that the model trained on high-resolution cells benefits implicitly from the 4.2M trained parameters in the SR model. To check whether a significantly larger PFlow model could have the same effect, we trained a variant with 4.6M parameters on the low-resolution cells. This model corresponds to the dashed line in Fig.~\ref{fig:pf_res}, enabling a direct comparison. While some improvement is indeed seen for kinematics, the same limitation in angular resolution persists.

When a particle interacts with calorimeter cells, it often deposits a significant portion of its energy in a single cell, making that cell ``hotter'' than the others. Consequently, the $\eta$ and $\phi$ values computed from the incidence matrix tend to gravitate toward those of the hottest cells. This results in the predicted particles not spreading evenly across the continuous space but instead forming diffused clumps at the midpoints of the highest resolution cells. In our case these are the cells in the ECAL1 layer. The effect is shown directly in Appendix~\ref{app:staircase}, where the predicted $\eta$ and $\phi$ follow a staircase as a function of their true values, with steps of one low-resolution cell. This clustering effect is the reason we observe the slightly bimodal structure in Figs.~\ref{fig:pf_res_eta} and \ref{fig:pf_res_phi} for the low-resolution model. Although the high-resolution model should theoretically display a similar bimodal pattern, because of its finer resolution, we do not see this effect in the figures.

\begin{figure*}[t]
\centering
\begin{subfigure}{0.36\textwidth}
  \centering
  \raisebox{0.48em}{%
    \includegraphics[width=0.95\linewidth]{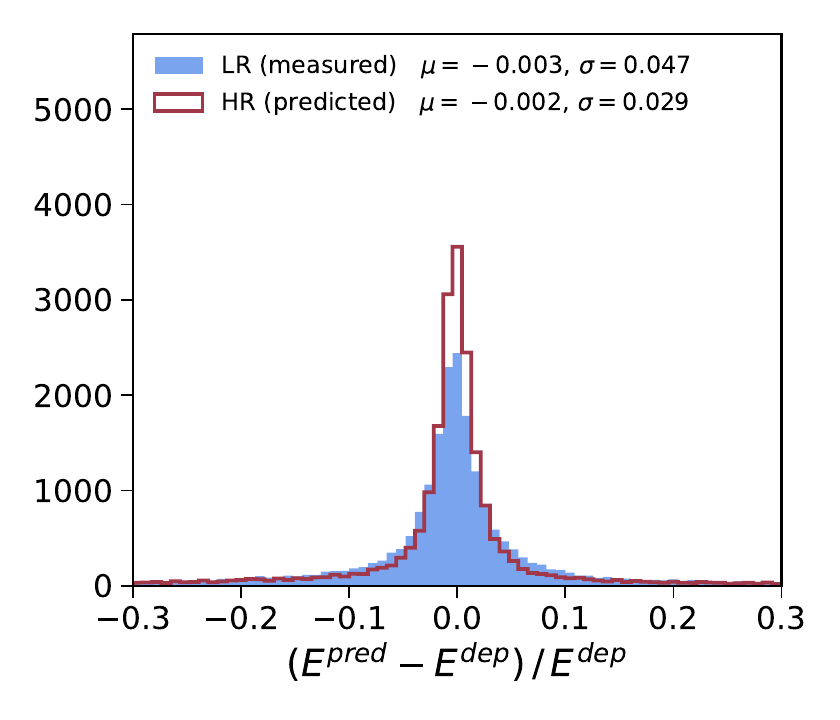}%
  }
  \caption{}
  \label{fig:pf_res_e_corr}
\end{subfigure}%
\hspace{5pt}
\begin{subfigure}{0.36\textwidth}
  \centering
  \includegraphics[width=0.95\linewidth]{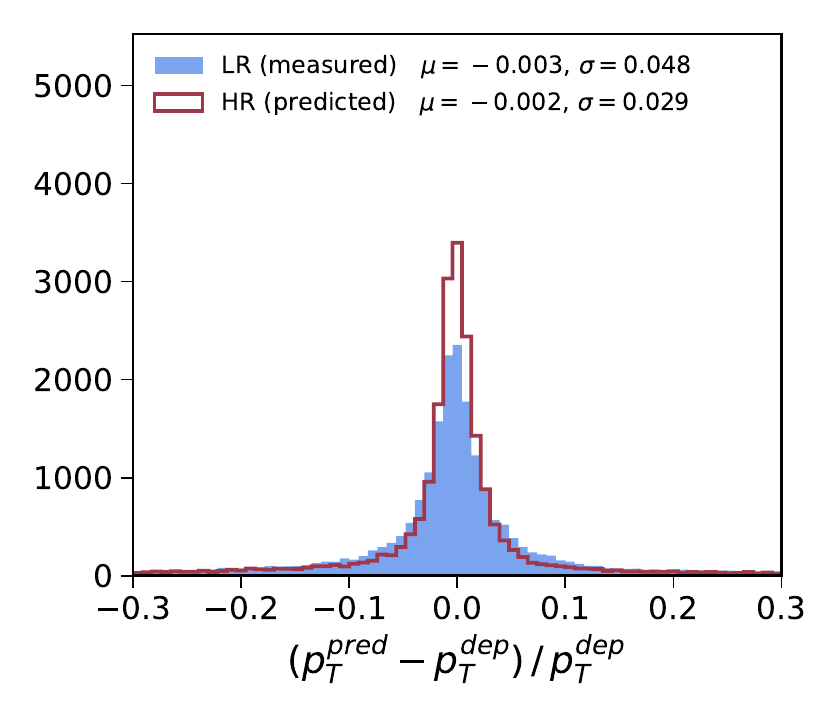}
  \caption{}
  \label{fig:pf_res_pt_corr}
\end{subfigure}

\begin{subfigure}{.36\textwidth}
  \centering
  \includegraphics[width=0.95\linewidth]{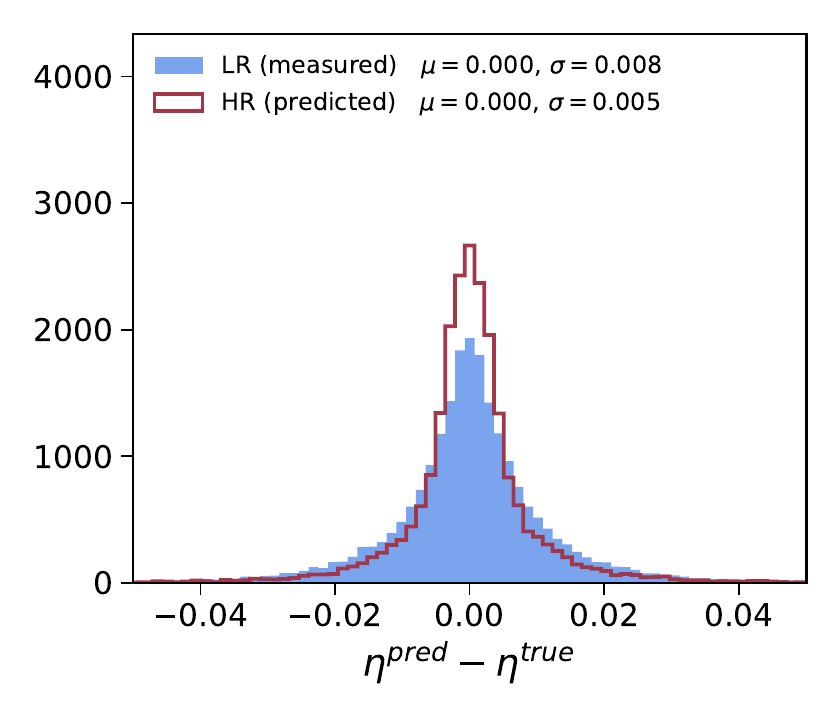}
  \caption{}
  \label{fig:pf_res_eta_corr}
\end{subfigure}%
\hspace{5pt}
\begin{subfigure}{.36\textwidth}
  \centering
  \includegraphics[width=0.95\linewidth]{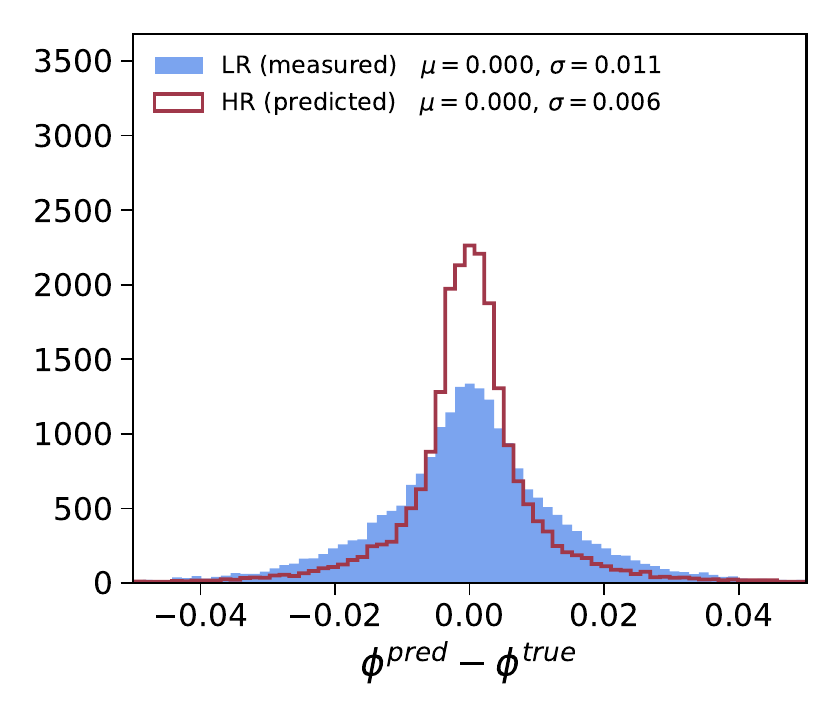}
  \caption{}
  \label{fig:pf_res_phi_corr}
\end{subfigure}
\caption{Particle-flow performance \textit{with the correction network}. (a) The relative energy residuals, (b) the relative transverse momentum residuals, (c) the pseudorapidity residuals, and (d) the azimuthal angle residuals between the truth particles and the reconstructed particles, for the PFlow models trained on measured low-resolution cells and on predicted high-resolution cells. The energy and transverse momentum residuals are computed with respect to the deposited energy, which is the quantity the models are trained to predict, and the corresponding transverse momentum. Both models are identical to those in Fig.~\ref{fig:pf_res} apart from the correction network applied to the proxy kinematics.}
\label{fig:pf_res_corr}
\end{figure*}

The structure described above is a property of the uncorrected estimator rather than of the input granularity, and it is removed once the correction network is included. Figure~\ref{fig:pf_res_corr} shows the same residuals for the two models with the correction network. The low-resolution model improves substantially and the bimodal structure disappears, reaching an angular resolution of $\sigma(\Delta\eta) = 0.008$. This is well below the value of $0.027$ expected from the cell size alone, which confirms that the correction network recovers subcell information from the energy sharing between neighboring cells. The high-resolution model also improves, and the difference between the two is reduced but does not vanish. The model trained on predicted high-resolution cells retains a factor of $1.5$ in $\eta$, $1.9$ in $\phi$ and $1.6$ in energy. Since both models can now access subcell information, the remaining difference is attributable to the input cells rather than to the form of the estimator. The two comparisons answer different questions: the first quantifies what superresolution offers to a fixed reconstruction, and the second what it offers when the reconstruction is itself free to correct for the granularity of its input.

Superresolution both suppresses noise and increases granularity, and the comparison above does not separate the two. To disentangle them, we sum the predicted high-resolution cells within each parent low-resolution cell. These cells carry the noise suppression of the superresolution model at the original granularity. A third particle-flow model with the correction network was trained on them, and its residuals are shown in Fig.~\ref{fig:denoise} in Appendix~\ref{app:sr_denoise_split}. The denoised cells recover most of the improvement in energy but only a small part of the improvement in $\eta$. The model trained on the predicted high-resolution cells still retains a factor of 1.4 in $\eta$, 1.5 in $\phi$ and 1.2 in energy over them. This split follows from the observables. The energy of a particle is a sum over cells and is unchanged by the downsampling that defines the low-resolution cells, so it can only be improved by noise suppression. The angular coordinates depend on how the energy is shared among cells, and for these the finer segmentation carries most of the gain. The denoised cells are derived from the superresolution model rather than from a denoiser trained at low resolution, so this separation is indicative rather than exact.

\begin{figure*}[t]
  \centering
  \includegraphics[width=0.85\textwidth]{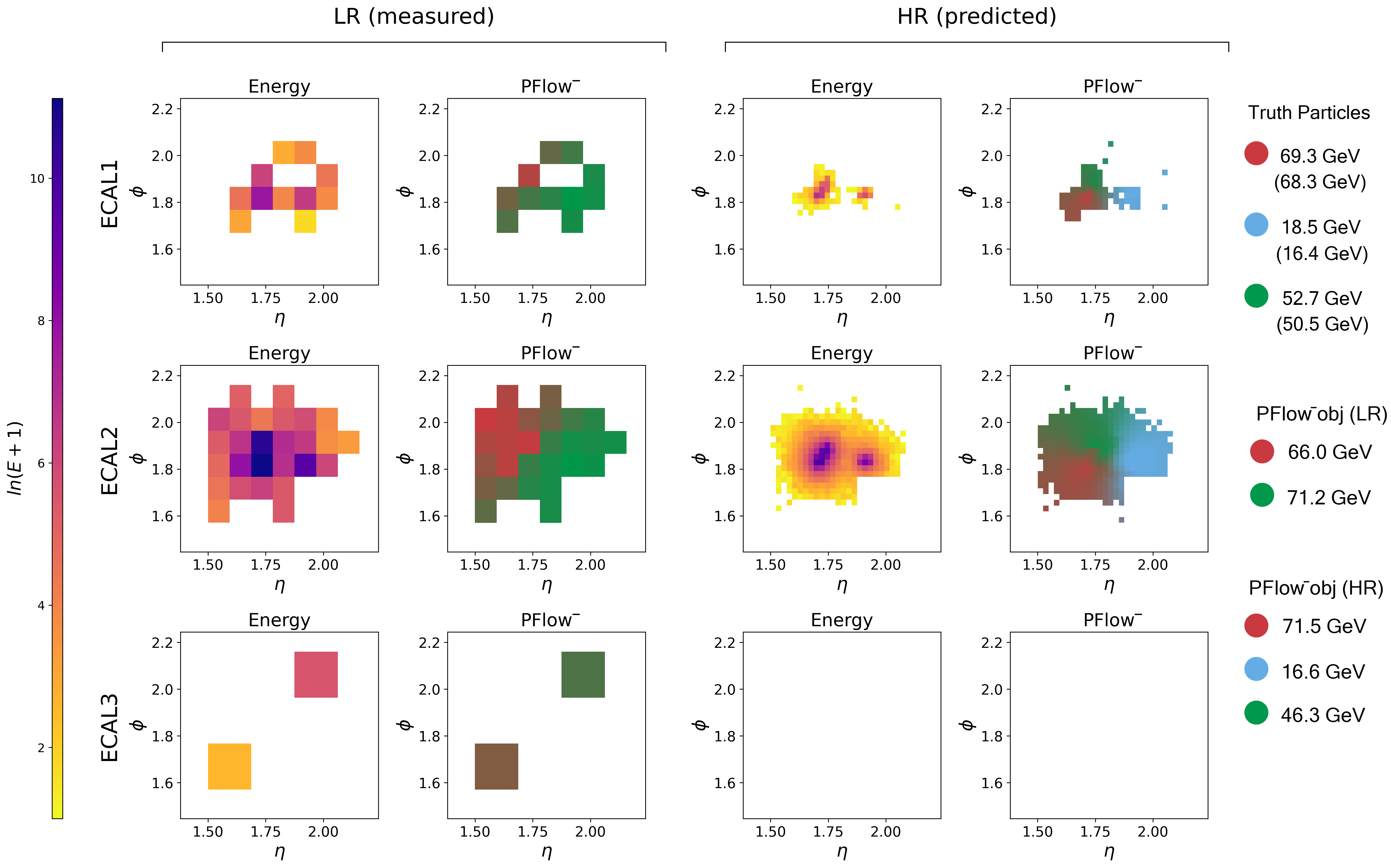}
  \caption{Event display illustrating the application of superresolution on particle flow, as well as the interpretability of the particle reconstruction model. Each cell is assigned a fractional weight based on the predicted attention or incidence values, which determine the cell colors. The legend shows the truth and predicted particles with their energies. The true deposited energy for the truth particles is shown in parentheses. The particle-flow model trained on the predicted high-resolution cells correctly identifies all three particles with very accurate energy estimations, while the model trained on low-resolution measured cells fails to disentangle the event into three particles and predicts only two particles.}
  \label{fig:pf_ed}
\end{figure*}

Figure~\ref{fig:pf_ed} presents a detailed event display showcasing the particle-flow model in action. In this case, the cardinality predictor is run first, and its output is fed into the kinematic predictor. Although the entire process occurs at the graph level, for clarity, we present the results in the $\eta-\phi$ plane as images.

\section{Discussion} \label{sec:discussion}
The results from the studies above indicate that superresolution is a viable and promising approach for improving reconstruction. In the following paragraphs, we address the question of how these improvements should be understood, discuss the path toward deployment in real experiments, and present areas of future work.

\subsection{Superresolution or hallucination}

A central question in this context is understanding the origin of the ``extra'' information that enhances low-resolution inputs into high-resolution outputs. The key insight lies in the patterns learned from the training data, particularly the nature of electromagnetic (EM) showers. Although stochastic on a case-by-case basis, the showers sample from a statistical prior $p(\mathrm{HR})$ that the network can learn from the high-resolution training examples. The model generates high-resolution features from the conditional distribution $p(\mathrm{HR} | \mathrm{LR})$, where the mapping is inherently one to many due to the stochastic nature of EM showers. While the model can approximate the manifold of consistent solutions arbitrarily well, the data processing inequality imposes a fundamental ceiling on its ability to recover the specific ``truth'' instance lost during measurement. Extra features thus stem from the information stored in model parameters, yielding samples that are distinct from the truth but physically consistent with its underlying prior. The same ceiling implies that superresolution cannot improve upon an optimal reconstruction from the low-resolution cells, and we do not claim that it does. What it provides instead is a representation: the subcell information already present in the low-resolution cells is expressed in the standard high-granularity form, where algorithms developed for fine-granularity detectors can use it directly. The relevant question for a reconstruction chain is what a given downstream model achieves in practice, and the comparison in Fig.~\ref{fig:pf_res_corr} shows that the improvement persists when both models are free to correct for the granularity of their input.

\subsection{Model dependence}

Training on EM showers from Standard Model particles will not equip the SR model to handle showers drawn from an anomalous prior. Examples include hypothetical particles with long lifetimes, exotic couplings, fractional electric charge, or magnetic monopoles. Such scenarios present a challenge to standard reconstruction techniques as well as SR which are tuned to Standard Model signatures. These limitations are a reminder that SR cannot be considered a replacement for actual high-granularity calorimeters.

\subsection{Data-simulation alignment}

A related corollary is that training on simulation can lead to degraded or unexpected performance on real data. This alignment problem affects machine learning models in general and has also been addressed in collider physics applications~\cite{ATLAS:2025rbr, CMS:2025kgf}. To mitigate the effect, future applications of this technique must include rigorous validation strategies to quantify this bias. While direct comparisons are difficult with experimental data, it is possible to develop small, high-granularity prototype modules to validate the network's outputs. In addition, techniques such as data augmentation can be employed to help reduce the model's dependence on features specific to the simulation~\cite{Zhong2025}.

\subsection{Superresolution as an auxiliary task}

Moreover, SR should be considered not just as an independent task but as an auxiliary to the broader reconstruction processes. By guiding the network to focus on relevant features, superresolution can complement traditional reconstruction algorithms, which explains why LR PFlow performance lags behind HR PFlow. In this paper, we explored SR as a preprocessing task, but it could also be embedded as a component of the PFlow training objective. The integration of SR techniques in this way has the potential to inform future detector designs and enhance the development of reconstruction algorithms, ultimately leading to more accurate and reliable data interpretation in collider experiments.

\subsection{Toward deployment}

To demonstrate a proof of concept, we have restricted our study to a simplified physics and detector setup. Several developments are needed to apply the SR technique to real experiments. First, the model needs to be trained on both hadronic and EM showers. Incorporating hadronic showers will introduce greater complexity due to the increased fluctuations inherent in these processes. Progressing toward full, realistic collision events will entail a significant rise in particle multiplicity due to jets and pileup. Including tracks as an input to the SR model may aid its performance in highly dense environments. To manage the model's memory demands, limiting graph connectivity and possibly clustering in angular space will be advantageous. Another important effect we have ignored presently is that of a magnetic field which alters the incident angle of particles and, in some detectors, their daughter particles.

\subsection{Ideas for future work}

While we have focused on superresolution in the angular space, it would be interesting to try this in the longitudinal direction, where the number of calorimeter layers is upscaled. Uncertainty quantification is another idea that could improve the interpretability of the SR outputs and flag instances of poor generation. Uncertainty quantification for generative models is an active area of research, with recent proposals targeting Bayesian inference for diffusion and flow models~\cite{kou2024bayesdiff}. Finally, while we have focused on generic particle reconstruction, future work could explore the potential of SR to enhance unfolding in specific topologies~\cite{Huetsch:2024quz}.

\begin{figure*}[t]
  \centering
  \includegraphics[width=0.75\textwidth]{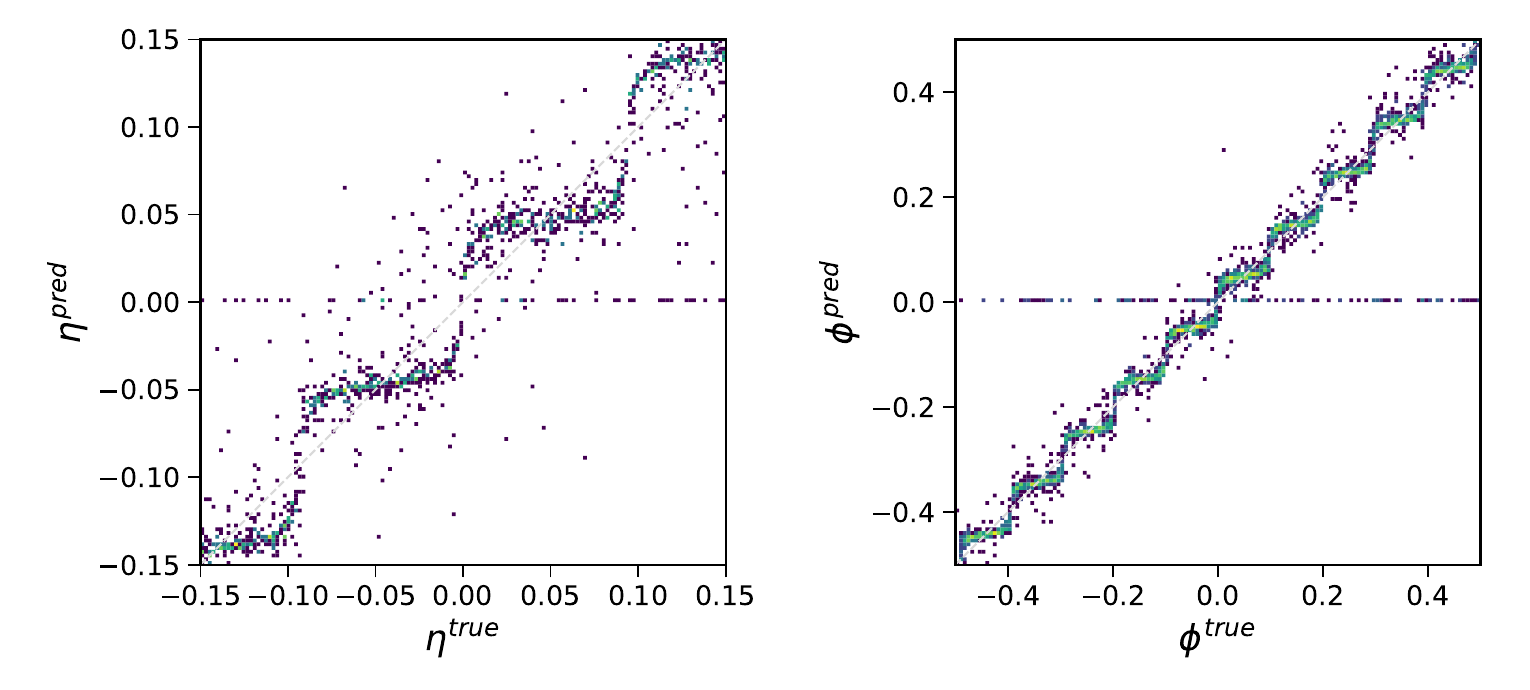}
  \caption{Predicted against true pseudorapidity (left) and azimuthal angle (right) for the particle-flow model trained on the measured low-resolution cells, without the correction network. The predictions are quantized in steps of one low-resolution cell. The dashed line indicates equality.}
  \label{fig:pf_staircase}
\end{figure*}

\section{Conclusion} \label{sec:summary}
In this work, we demonstrated, for the first time, that superresolution techniques can be effectively incorporated into a simplified LHC-like particle-flow reconstruction chain. We introduced a novel graph-based model for superresolution that leverages the powerful flow-matching framework alongside a transparent and interpretable particle-flow model. However, the key takeaway is not the specific models employed; rather, it is that superresolution can indeed enhance reconstruction. Our approach faithfully replicates high-resolution cells event by event, leading to significantly improved modeling of cell substructure variables. Moreover, utilizing these high-resolution cells for particle flow substantially enhances both particle cardinality and angular resolution. This improvement persists when both particle-flow models are equipped with a correction network for the proxy kinematics, showing that it does not follow from the restricted form of the default kinematic estimator. We believe these findings motivate further exploration of SR in reconstruction pipelines. Detector designs are evaluated through the reconstruction performance they deliver in simulation, so a software step that improves that performance at fixed granularity bears directly on the performance-to-cost trade-off, without standing in for the measurement a finer detector would make.

\begin{acknowledgments}
We thank Dmitrii Kobilianskii for fruitful discussions on flow-matching implementation and Kyle Cranmer and Yihui Ren for their valuable insights. E.\,G. also acknowledges support from ISF Grant No.~2871/19, the BSF-NSF Grant No.~2020780, and the Weizmann Institute of Science and Mohamed bin Zayed University of Artificial Intelligence (MBZUAI) collaboration grant. We also thank the Benoziyo Center for High Energy Physics for supporting our research.
\end{acknowledgments}

\section*{Data Availability}

To facilitate reproducibility, the code used for the analyses presented in this paper is available on GitHub~\cite{sr_code}. The dataset generated and analyzed during this study can be accessed through Zenodo~\cite{sr_data}.

\appendix

\section{QUANTIZATION OF THE PROXY KINEMATICS}
\label{app:staircase}

The proxy kinematics are obtained as incidence weighted averages of fixed cell coordinates, so the predicted position of a particle is a weighted mean of the centers of the cells that receive its energy. When most of the energy is collected by a single cell, the prediction is drawn toward the center of that cell. Figure~\ref{fig:pf_staircase} shows the predicted $\eta$ and $\phi$ against their true values for the model trained on the low-resolution cells. The predictions follow a staircase whose steps correspond to the width of a low-resolution cell in the first two ECAL layers, $0.094$ in $\eta$ and $0.098$ in $\phi$. This is the origin of the bimodal structure seen in Figs.~\ref{fig:pf_res_eta} and \ref{fig:pf_res_phi}. The effect follows from the form of the estimator rather than from the input cells, and it is absent once the correction network is included.

\clearpage
\onecolumngrid
\section{SEPARATING DENOISING FROM GRANULARITY}
\label{app:sr_denoise_split}

\begin{figure}[h]
  \centering
  \includegraphics[width=0.8\textwidth]{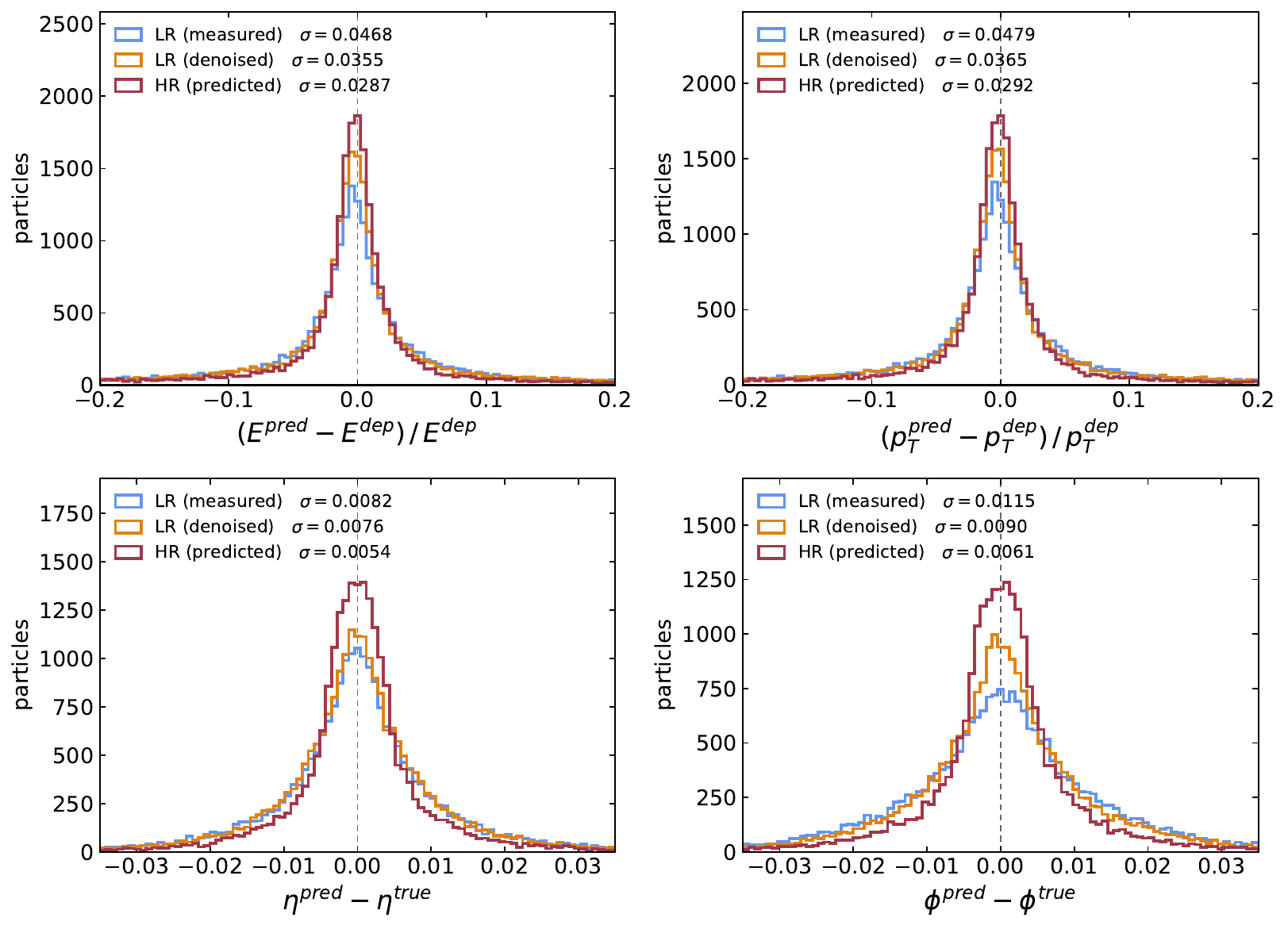}
  \caption{Particle-flow performance with the correction network for three sets of input cells: (top left) the relative energy residuals, (top right) the relative transverse momentum residuals, (bottom left) the pseudorapidity residuals, and (bottom right) the azimuthal angle residuals. The three models are identical apart from their inputs. The denoised low-resolution cells are obtained by summing the predicted high-resolution cells within each parent low-resolution cell, so they carry the noise suppression of the superresolution model at the original granularity, and the difference between them and the predicted high-resolution cells reflects the granularity of the representation alone.}
  \label{fig:denoise}
\end{figure}
\twocolumngrid

\bibliography{super_res}

\end{document}